\documentclass
{mn2e}
\usepackage{epsfig}
\usepackage{amsmath, amssymb,bm}

\title[Accretion disc viscosity and warps]{Accretion disc viscosity:
  what do warped discs tell us?}

\author[A.R King et al.]{A.R. King$^1$, M. Livio$^2$, S.H. Lubow$^2$ \&
  J. E. Pringle$^{1, 2, 3}$ \thanks{E-mail: jep@ast.cam.ac.uk}\\ 
1. Theoretical Astrophysics Group, University of Leicester, Leicester
LE1 7RH, UK\\
2. Space Telescope
  Science Institute, 3700 San Martin Drive, Baltimore, MD 21218, USA\\
3. Institute of Astronomy, University of Cambridge, Madingley Rd,
Cambridge CB3 OHA, UK.}

\begin{document}

\pagerange{\pageref{firstpage}--\pageref{lastpage}} 
\pubyear{2012}
\maketitle

\label{firstpage}

\begin{abstract}

Standard, planar accretion discs operate through a dissipative
mechanism, usually thought to be turbulent, and often modelled as a
viscosity. This acts to take energy from the radial shear, enabling
the flow of mass and angular momentum in the radial direction. In a
previous paper we discussed observational evidence for the magnitude
of this viscosity, and pointed out discrepancies between these values
and those obtained in numerical simulations. In this paper we discuss
the observational evidence for the magnitude of the dissipative
effects which act in non--planar discs, both to transfer and to
eliminate the non--planarity. Estimates based on the model by Ogilvie
(1999), which assumes a small--scale, isotropic viscosity, give
alignment timescales for fully ionized discs which are apparently too
short by a factor of a few compared with observations, although we
emphasise that more detailed computations as well as tighter
observational contraints are required to verify this conclusion. For
discs with low temperature and conductivity, we find that the
timescales for disc alignment based on isotropic viscosity are too
short by around two orders of magnitude. This large discrepancy
suggests that our understanding of viscosity in quiescent discs is
currently inadequate.

\end{abstract}


\section{Introduction}

%Viscous accretion discs have loosely two viscosities, the normal
%$\nu$ and $\nu_2$, or equivalently the usual $\alpha$ and also
%$\alpha_2$.

Accretion discs have major importance for many branches of
astrophysics. The fundamental process determining how they work is
usually called viscosity. This determines the transport of mass and
angular momentum within the disc, and as a result how much energy this
releases. Despite considerable efforts, the physical origin of the
viscosity is still at best only partially understood. This makes it
worthwhile to consider what we can learn about it from the
behaviour of observed disc systems. We attempt this here.

The simplest case of a plane disc has been extensively studied, so we
review it only briefly before considering cases where the disc is
tilted or warped, which is our main purpose in this paper. In this simplest
case a geometrically thin Keplerian disc lies in a plane ($z = 0$ in
cylindrical polar coordinates $R, \phi, z$) and the $R\phi$ stress
transmitted by viscosity transports angular momentum outwards and mass
inwards (Pringle 1981). The conservation laws for these two quantities
combine to give an equation governing the radial motion of the surface
density $\Sigma(R, t)$,
%%%%%%%%%%%%
%%(see \ref{diffus} below). 
%%%%%%%%%%%%%%%%
which shows that the surface density diffuses inwards on a timescale
\begin{equation}
t_{\rm visc} \simeq {R^2\over \nu}  
\label{tvisc}
\end{equation}
at disc radius $R$, where $\nu$ is the viscosity. 
For a disc semi-thickness $H \ll R$ we follow
Shakura \& Sunyaev (1973) and define the dimensionless parameter $\alpha$ as
a measure of the vertically averaged viscosity through the relation
\begin{equation}
\label{defalpha}
\nu = \alpha H c_s.
\end{equation}
Here $c_s$ is a vertically averaged measure of the disc sound speed,
related to $H$ by hydrostatic equilibrium in the $z$-direction as
\begin{equation}
\label{CS}
c_s = H \Omega.
\end{equation}
Shakura \& Sunyaev (1973) comment that if the viscous mechanism
is either hydrodynamic or magneto-hydrodynamic turbulence, then rapid
dissipation of such turbulence at supersonic or super-Alfv\'enic
speeds is likely to ensure that $\alpha \le 1$.

If the system is time--variable, as in dwarf novae or soft X--ray
transients, we get an estimate of $\nu$ or $\alpha$ by comparing the
result (\ref{tvisc}) with the observed timescales of variability and
estimates of the disc radius. For outbursts of both types of system,
where the disc is fully ionized and gas pressure dominates, this
suggests that $\alpha$ is in the range $0.1 \le \alpha \le 0.3 $
(King, Pringle \& Livio, 2007; Kotko \& Lasota, 2012).

In the rest of this paper we ask what we can learn from observations
of another aspect of disc viscosity, namely the $Rz$ stress involved
in possible tilting and warping of the disc.

\section{Propagation of disc warps}

The response of gaseous discs to tilts or warps depends crucially on
whether the viscosity is large or small, in the sense that it depends
crucially on whether the radial communication of the tilt is mainly
by pressure forces, or alternatively by viscous stress.

For high viscosity, i.e. $\alpha \gg H/R$, the warp is transmitted
radially by viscous stresses. Here too, by consideration of
conservation of mass and of angular momentum, for matter orbiting
locally in a simple disc--like geometry, it is possible to obtain
idealised equations for the evolution of the disc (both surface
density and tilt), involving two viscous parameters: the usual $\nu$
relating to the $R\phi$-stress, and a second viscosity $\nu_2$ relating
to the $Rz$-stress (Papaloizou \& Pringle 1983, Pringle 1992).  If we
imagine each disc annulus to have a normal unit vector ${\bf l}(R,t)$,
then for small disc tilt angle $\beta(R,t) \ll
1$, we have
\begin{equation}
{\bf l} = (\beta \cos \gamma, \beta \sin \gamma, 0)
\end{equation}
in Cartesian coordinates, with $OZ$ perpendicular to the disc, and
where $\gamma(R,t)$ represents the azimuth of the disc tilt at radius
$R$ and time $t$. Then if we write $W \equiv \beta e^{i \gamma}$
(Petterson 1997a) and ignore external torques, for a steady disc (one
for which $\Sigma$ is independent of time) the warp evolves as
\begin{equation}
\label{Wdisc}
\frac{\partial W}{\partial t} = \frac{\nu_2}{2R}
\frac{\partial}{\partial R} \left( R \frac{\partial W}{\partial R}
\right).
\end{equation}
So in these circumstances, warp propagation obeys a diffusion
equation and the timescale on which the warp is smoothed out is
approximately
\begin{equation}
\label{tnu2}
t_{\rm damp} = t_{\nu_2} \sim \frac{R^2}{\nu_2}.
\end{equation}

Ogilvie (1999, 2000) has considered the particular case in which the
disc viscosity can be assumed to be small-scale (i.e. mean free path
much less than $H$) and isotropic, so that the disc fluid obeys the
usual Navier-Stokes equations. He finds that the simplified equations
given by Pringle (1992) provide an adequate description of the disc
evolution, and there is an additional, lower order effect which gives
rise to precession of the disc annuli. He also shows for small
disc tilts (see also Lodato \& Price, 2010) that
\begin{equation}
\frac{\nu_2}{\nu} = \frac{1}{2 \alpha^2} \frac{4(1 + 7 \alpha^2)}{4 + \alpha^2}.
\end{equation}
For small values of $\alpha$ (but still with $\alpha >> H/R$) this gives 
\begin{equation}
\label{alpha2}
\frac{\nu_2}{\nu} \approx \frac{1}{2 \alpha^2}.
\end{equation}
This result is somewhat counter-intuitive, since it indicates that as
$\alpha$ decreases, $\nu_2$ increases.The relationship comes about for
Keplerian discs, for which the epicyclic frequency and orbital
frequency almost coincide, so that a slight disc tilt produces a
resonant response in internal horizontal disc motions (Papaloizou \&
Pringle, 1983; see the discussion in Lodato \& Pringle 2007).

%Discuss briefly here the theoretical ideas about the relationship
%between $\nu$ and $\nu_2$. Papaloizou \& Pringle 1983; Ogilvie 1999,
%2000. See physical explanation of why $\nu_2/\nu \sim 1/\alpha^2$ in
%Lodato \& Pringle 2007.

For the opposite case of low viscosity, that is $\alpha \ll H/R$,
the warp is communicated radially by pressure. A small warp travels as
a linear wave with wave speed $v_w = (1/2) c_s$, where $c_s$ is (an
appropriate vertical average of) the local sound speed in the disc
(Papaloizou \& Lin, 1985; Pringle, 1999). In this case the warp
produces resonant horizontal shearing motions within the disc which
are then subject to damping by the internal disc viscosity. Warp
damping takes place over approximately $1/\alpha$ orbital periods of the
disc so that locally
\begin{equation}
t_{\rm damp} \sim \frac{1}{\alpha \Omega} .
\end{equation}

%Note: this is what appears to be given in Lubow et al 2002, Nixon \&
%Pringle 2010, and found in the simulations but it needs more
%complicated analysis in order to compute the disc alignment timescale
%given in Bate et al 2000; Lubow \& Ogilvie 2000.

%We need to give here the realignment timescale for a disc in a binary
%in the low alpha case, so that we can refer to it below.

%In this paper we investigate the extent to which limits can be put on
%the damping of disc warps, and so on the effective second disc
%viscosity, from observational considerations.

\section{Systems with misaligned discs}

In considering the evolution of misaligned discs, we first need to
identify astrophysical systems in which the disc is thought to be
misaligned. The most obvious candidates in this regard are
semi--detached binary systems. In some of these the accretion disc
around the mass--gaining star (here the primary) is for some reason
misaligned with the binary orbit. It seems very likely that this
misalignment results in some way from the presence of a binary
companion orbiting outside the disc, although there is as yet no fully 
worked--out theory of this. In particular, although mass transfer from
the companion may be an ingredient of the mechanism causing the
misalignment, it cannot of itself be the direct cause: since
mass transfer in the normal way supplies only an aligned component of
angular momentum, it actually tends to {\it reduce} misalignment.

There are two main types of semi--detached binary system in which the
accretion disc is thought to be misaligned -- X-ray binaries which
display long term `superorbital' periods or modulations (summarised by
Kotze \& Charles 2012), and cataclysmic variables displaying
so--called `negative superhumps' (summarised by Olech et al 2009). We
discuss one example from each class, choosing the one which
seems to give the most information about the viscosity in the system.

In each case the most readily observable quantity is the timescale on
which the disc aligns with the orbital plane once the force driving
the misalignment has been removed. To see what this tells us about the
behaviour of viscosity in a warped disc we estimate the timescale for
discs to align solely under the effect of viscosity. The realignment
of the disc with the orbital plane then comes about through tidal
torques which dominate at the outer disc edge. These tides cause
differential precession of the outer disc elements. The resulting
twist in the disc is then smoothed out by viscous effects, causing
alignment. This viscous alignment timescale overestimates the true
alignment timescale if significant mass transfer continues, since this
adds aligned angular momentum to the disc, as noted above.

\subsection{Alignment timescale for tidally truncated discs}

If the disc radius is $R_d$, the orbital timescale is $\Omega = (G
M_1/R_d^3)^{1/2}$, where $M_1$ is the mass of the disc's central star
(the primary). We write $1/\omega_p$ as the timescale for (retrograde)
precession induced by the tidal effect of the secondary star, so that
the precession period is $P_p = 2 \pi/\omega_p$.

\subsubsection{High viscosity, $\alpha > H/R$}

In this case, the warp is transmitted radially primarily by viscous
stresses. Here we are concerned with what happens in the regime
$t_{\nu_2} \omega_p < 1$, when the viscous stresses are able to make
the disc act as a cohesive whole (when this is not the case,
differential precession is able to tear the disc into separate rings;
Nixon et al. 2012).

For a steady disc subject to an external precessional torque the warp
equation(\ref{Wdisc}) can be written schematically as
\begin{equation}
\frac{\partial W}{\partial t} = i \omega_p W + \frac{1}{2} \nu_2
\frac{1}{R} \frac{\partial}{\partial R} \left( R \frac{\partial
  W}{\partial R} \right).
\end{equation}

Multiplying this equation by $W^\dagger$, the complex conjugate of
$W$, and then adding the resulting equation to its complex conjugate
we get
\begin{equation}
\frac{\partial}{\partial t} |W|^2 = \frac{1}{2} \nu_2 \left[
  \frac{W^\dagger}{R} \left\{ \frac{\partial}{\partial R} \left( R
  \frac{\partial W}{\partial R} \right) \right\} + \frac{W}{R} \left\{
  \frac{\partial}{\partial R} \left( R \frac{\partial
    W^\dagger}{\partial R} \right) \right\} \right].
\end{equation}

We take the integral of this over the disc, integrating the
r.h.s. by parts and assuming zero torque boundary conditions. This gives
\begin{equation}
\frac{\partial }{\partial t} \int |W|^2 \,R dR = - \frac{1}{2} \nu_2
\int \left| \frac{\partial W}{\partial R} \right|^2 \, R dR.
\end{equation}

We can make estimates for each side of this equation. The l.h.s. is
simply $\sim W^2/t_{\rm align}$. To estimate the r.h.s., we note that if
$t_{\nu_2} \sim 1/\omega_p$ the viscous stresses acting to flatten the
disc act on the same timescale as the precessional torques which tend
to twist it up, and so we might expect that $\partial W/ \partial R
\sim W/R$. As the viscous timescale decreases, precession becomes less
able to twist the disc, and so we can estimate for $t_{\nu_2} \omega_p
< 1$ that
\begin{equation}
\frac{\partial W}{\partial R} \sim \frac{W}{R} \times (t_{\nu_2} \omega_p).
\end{equation}

Using these estimates, together with equations(\ref{defalpha}),
(\ref{CS}) and (\ref{alpha2}) our estimate becomes
\begin{equation}
\label{talignvisc}
t_{\rm align} \approx \omega_p^{-1} \alpha^{-1} (H/R)^2 (\Omega_d/\omega_p),
\end{equation}
where all quantities are evaluated at $R_d$.

\subsubsection{Low viscosity, $\alpha < H/R$}

The timescale for disc alignment in this case is considered by
Bate et al. (2000) and by Lubow \& Ogilvie (2000). In the low
viscosity limit the warp propagates as a linear wave at a speed $v_w =
c_s/2$. Provided that
\begin{equation}
\omega_p \le {c_s\over R_d}
\end{equation}
the disc can precess as a whole (Papaloizou \& Terquem 1995, Larwood
et al 1996); otherwise the disc can be torn apart (Larwood et al.,
1996). This condition can also be written
\begin{equation}
\frac{\omega_p}{\Omega_d} \le \frac{H}{R}.
\end{equation}
For the systems we are concerned with, Bate et al (2000) estimate that
$\omega_p/\Omega_d \approx 0.01$ so that this inequality is easily
satisfied. Then Bate et al. (2000) estimate the timescale for the disc
to align with the binary orbit as
\begin{equation}
t_{\rm align} \approx \omega_p^{-1} \alpha^{-1} (H/R)^2 (\Omega_d/\omega_p)
\end{equation}
(all quantities evaluated at $R_d$) which is exactly the same formula
given in the high viscosity case.

\section{X-ray binaries: Hercules X--1/HZ Herculis}

The Hercules X--1/HZ Herculis system is a semi--detached binary system
in which an F star, mass $M_2 \approx 2 M_\odot$, transfers mass to a
neutron star, mass $M_1 \approx 1 M_\odot$. The binary period is $P =
1.7$ d. This period is apparent both in X--rays, because the neutron
star undergoes eclipses by the F star, and in the optical, because the
side of the F star facing the X-ray source is so strongly heated that
it appears more like an A star on that face. The system parameters are
quite well determined because the radial velocity of the neutron star
can be measured by observing the apparent change of its 1.24 s
rotation period around the orbit. In addition, the X-ray luminosity
undergoes a regular variation with a period of $P_{\rm long} \approx 35$
d. The 35~d X--ray flux curve for Her X--1 consists of two peaks,
equally separated in time, with a small residual flux in between. The
larger peak lasts about 10 days, and the smaller peak, occurring about
half a cycle later, lasts about 5 days (see the review by Priedhorsky
\& Holt, 1987). Throughout the 35~d period the heating of the face of
the companion star continues essentially unchanged, although there are
subtle changes in the optical light curve. These changes (both the
X-ray and the optical) led Gerend \& Boynton (1976) to propose a
detailed model for the system in which the accretion disc around the
neutron star is slightly tilted relative to the orbital plane and
precesses retrogradely with a period of 35~d. The model allows the
secondary star to be visible to the X--rays at all times during the
35~d cycle, but the illumination pattern on the face of the secondary
varies with the synodic period $(1/P + 1/P_{\rm long})^{-1}$ which is a
few per cent {\em less} than the orbital period (for retrograde
precession). Then, provided that the line of sight makes an angle to
the orbital plane which is non--zero but slightly larger than the
angle of disc tilt, the neutron star becomes directly visible
(and so the X--rays turn on) twice every precession period, with the
larger and longer peak occurring when the disc is tilted towards the
observer, and the smaller, shorter peak when the disc is tilted away.

The cause of the disc tilt is thought to be radiation warping (Petterson,
1977b,c; Pringle 1996). Modelling by Wijers \& Pringle 1999, and by
Ogilvie \& Dubus 2001 shows that this can work provided that the mass
input occurs at small radii (for example, Ogilvie \& Dubus assume that
the stream adds mass at the circularization radius). This ensures that
the disc shape is such that the radiation torques and the tidal
torques at the disc edge act in the same retrograde direction. From
their simulations Wijers \& Pringle find that the parameters of Her
X--1 can be fitted, assuming that $\nu_2/\nu = 1/2 \alpha^2$, with
$\alpha = 0.27$, as might be expected for a fully ionized disc (King
et al 2007). Ogilvie \& Dubus take $\alpha = 0.3$ and find that for
this value Her X--1 lies in that part of the parameter space where a
steadily tilted and precessing disc is expected (see also Kotze \&
Charles 2012). It is also worth remarking that, according to Ogilvie
\& Dubus (2001), Her X--1 lies close to the stability line, in that a
reduction of the luminosity by a small factor would result in
stability and an aligned disc.

So far this tells us little directly about $\nu_2$. But it does
  suggest that if the disc behaves as a Newtonian fluid (as in the
  theory of Ogilvie, 1999), the disc in Her X--1 has $\alpha$ in the
  range $0.1 \le \alpha \le 0.3$. This is the same range needed to
  explain dwarf nova decay timescales for discs with similar physical
  properties. This range also implies that for the disc in Her X--1,
  $\alpha \gg H/R \approx 0.04$ (Wijers \& Pringle, 1999) and so that
  warps propagate diffusively in this system, as we have assumed.

The most promising way of estimating $\nu_2$ is to find an
observational measure of the rate at which disc realignment occurs
when the forcing mechanism is turned off, or at least turned down. For
a disc in which the warp is propagated diffusively, the realignment
timescale is given by equation (\ref{talignvisc}). If the observed
precession is caused predominantly by tidal effects (cf. Larwood,
1998), we may assume that the dynamical driven precession
timescale is comparable to the observed precession timescale $P_{\rm
  long}$ (recall that the radiation torques which cause the disc tilt
also provide disc precession at some level) so that $P_p \approx
P_{\rm long} \approx 35$~d. For the parameters of the Her X--1/HZ Her
system, we take $q = M_2/M_1 = 2$, the ratio of disc radius to binary
separation $R_d/a = 0.24$, and the disc surface density as a power law
$\Sigma \propto R^{-3/2}$, so that from Bate et al (2000) we obtain
\begin{equation}
\frac{\omega_p}{\Omega_d} \approx 0.004.
\end{equation}

Then using equation (\ref{talignvisc}) and assuming that $\alpha \approx
0.3$, $H/R \approx 0.04$ and $\omega_p/\Omega_p \approx 0.004$, we
find that
\begin{equation}
t_{\rm align}(\rm Her  X-1) \sim 1.3 \omega_p^{-1} \sim 7 {\rm d}.
\end{equation}

From time to time the X--ray flux of Her X--1 drops and the system
enters what is known as an anomalous low state. This occurred in
1983/84 (Parmar et al 1985), in 1998/99 (Coburn et al 2000; Still et
al 2001) and in 2003/2004 (Jurua et al 2011, see also Kotze \& Charles
2012). Despite the X--ray flux drop, the heating of the companion
star HZ Her continues almost unchanged, indicating that although the
disc inclination or disc shape obscures the direct line of sight to
the neutron star, accretion continues and the face of much of the
companion star is still able to receive X--ray flux. As remarked
above,the 35 d X--ray flux curve for Her X--1 consists of two peaks. The
larger peak last about 10 days, and the smaller peak, occurring about
half a cycle later lasts about 5 days. The tilted disc models
suggest that during the large peak the disc is tilted towards the
observer, and during the small peak it is tilted away from the
observer (Gerend \& Boynton, 1976; Petterson, 1977c; Priedhorsky \&
Holt, 1987). Thus an anomalous low state can come about if the disc
becomes for some reason less inclined to the orbital plane, and thus
more aligned with the line of sight (Coburn et al 2000).

Thus it seems reasonable to conclude that the onset of an anomalous
low state is caused by the decrease in the disc inclination relative
to the orbital plane, caused presumably by some decrease in the
self--illumination of the disc and so a decrease in the physical cause
of the warp. In this regard we note that Thomas et al (1983) remark
that the amount of X--ray heating of HZ Her seen in the period 1979 --
1982, prior to the 1983 anomalous low state, seemed to have decreased
by around 15 per cent. The timescale on which this disc alignment
takes place can be judged by the timescale for the onset of the
anomalous low state. This appears (Coburn et al 2000, Still et al
2001) to occur on a timescale of order $P_p$.

In summary, although our estimated disc alignment timescale is
slightly shorter than observations indicate, within the limits of
the approximations made here, the whole picture fits reasonably well
with a model in which warped disc motions are treated as those of a
Newtonian fluid, subject to a Navier--Stokes viscosity, with $\alpha
\approx 0.3$.

\section{Dwarf novae: V503 Cygni}

The SU UMa subclass of dwarf novae have short orbital periods, in the
range 1.3 -- 2~h, and display two types of outburst: normal
outbursts and `superoutbursts' (e.g. Warner, 1995). The
latter occur less frequently, but are brighter and last longer. During
the superoutbursts a photometric modulation, with period longer than
the orbital period by a few per cent (dependent on mass ratio $q =
M_2/M_1$), is some times apparent. This modulation is known as a
`superhump', and is thought to be caused by the disc becoming large
enough to be driven eccentric by a 3:1 resonance (Lubow 1991a, 1991b),
and precessing in a prograde direction. Some of these systems, of
which a clear cut example we consider here is V503 Cyg (Harvey et al
1995), display a photometric modulation during superoutburst which has
a period slightly {\em shorter} than the orbital period, again by a few
percent. This would then correspond to a disc precessing in the
retrograde direction and is known as a `negative superhump'. The basic
model for the origin of negative superhumps is outined by Wood and
Burke (2007). They suggest that the disc is tilted with respect to the
orbital plane, resulting in retrograde tidal precession, and that the
modulation is caused by the accretion spot, where the mass transfer
stream strikes the disc, varying in distance from the primary white
dwarf, and thus varying in brightness. The question here is what
causes the disc tilt, and under what circumstances such a tilt can be
maintained.

For dwarf novae, the radiation warping discussed above fails by many
order of magnitude, as indeed it does for many of the low mass X--ray
binaries (Ogilvie \& Dubus 2001). In the absence of this, the major
problem with any warping mechanism is that if the mass transfer stream
remains in the orbital plane, so that the transferred matter has no
component of angular momentum that lies in the orbital plane, the
resultant disc must also lie in the orbital plane. What is required is
a mechanism which imparts to the disc a net component of angular
momentum that lies in the orbital plane. Currently the only viable
mechanism proposed in the literature is the suggestion by Smak (2009)
that the mass transfer stream through the inner Lagrange point $L_1$
has a component perpendicular to the disc plane which oscillates in
phase with the binary period. He suggests that this comes about
because the tilted disc enables the neighbourhood of the $L_1$ point
to be heated in an asymmetric manner, and one which varies on the
orbital period. It is worth remarking here that modulation of mass transfer
caused by irradiation has been discussed by various authors
(e.g. Arons, 1973; Basko \& Sunyaev, 1973; Viallet \& Hameury, 2007)
and that a similar suggestion about diverting the mass transfer stream
out of the orbital plane was made by Shakura et al (1999) in
order to account for some of the dips seen in the X--ray light curve
of Her X--1.

This mechanism can only work when the illumination of the surface of
the secondary is enough to influence the mass transfer process and so
is likely to occur only during outburst and/or superoutburst. The
preferential occurrence of negative superhumps in SU UMa systems
(along with some novalikes) indicates that we may need superoutbursts
where the mass transfer rate is higher, but that this might also occur
because the secondaries in these systems have low mass and so low
intrinsic surface brightness.  Smak (2004: Fig. 1) plots the ratio of
irradiation flux to intrinsic stellar flux at a point close to $L_1$
for a number of (un--named) dwarf novae. It is notable that in the
systems with shorter periods, i.e. those which show superoutbursts,
the ratio is much larger (in the range 150 -- 200) compared to the
longer period systems (all $< 100$, and typically $ < 20$).

\subsection{Basic properties of V503 Cyg.}

The properties of V530 Cygni are discussed by Harvey et al
(1995). The orbital period, derived from radial velocities, is P =
111.9 min = 0.078 d.~\footnote{Note that in many `negative superhump'
  systems the orbital velocity has to be identified from among
  several photometric periods apparently present in the data.} The
negative superhump has period $P_- = 109.0$ min, which implies
precession period of $P_p = 4174$ min = 2.90 d = 37.2 P.

Dwarf nova outbursts are though to be caused by limit cycle behaviour,
with the disc jumping between a state in which the disc is hot and
fully ionized, and the viscosity and accretion rate are high, and a
state which is cool, not fully ionized, and the viscosity and
accretion rate are low (see e.g. Lasota, 2001). Models of dwarf nova
outbursts (e.g. Cannizzo, 1994; Hameury et al 1998) typically
have

outburst: $H/R \approx 0.02 - 0.03$ and $\alpha_h = 0.1 - 0.3$.

quiescence: $H/R \approx 0.006$ and $\alpha_c = 0.01 - 0.02$

Note that in both cases $\alpha > H/R$, and therefore we expect the
warp to evolve viscously at all times. But we should also note that
the limits on $\alpha$ in the low state are probably not
well--defined. That is, although we do require $\alpha_c \ll \alpha_h$
in order to get outbursts of the right magnitude, it is possible that
$\alpha_c$ could be a lot smaller than in these
models. One method of trying to identify $\alpha_c$ is to model the
variation of disc radius through the outburst cycle (cf. Ichikawa \&
Osaki 1992). The smaller $\alpha_c$, the less the disc evolves in
quiescence and so the smaller the quiescent disc. In this context we
note the finding by Smak (1999) that discs in quiescence appear to be
smaller than models predict and to lie well away from the outer disc
tidal truncation radius. Therefore in the following analysis we
consider $\alpha_c$ to be relatively ill--determined and see to what
extent we can set independent limits.

The observations we are trying to account for are described by Harvey et
al. (1995), and relate to the outburst behaviour of V503 Cyg in
1994. At that time the superoutbursts occurred every 88~d, lasting
$\approx 10$ d, and normal outbursts occurred every $30 \pm 3$ d,
lasting $\approx 3$ d. Negative superhumps were seen throughout
all stages of the eruption cycle (both in outburst and in quiescence)
at approximately constant amplitude.

\subsection{Timescale estimates}

Suppose that the the masses in V503 Cyg take typical values of $M_2 =
0.2 M_\odot$, $M_1 = 0.8 M_\odot$, so that $q = 0.25$ and the total
mass $M = 1 M_\odot$. Then with the binary period of 112 min we have a
binary separation of
\begin{equation}
a = 5.34 \times 10^{10}  {\rm cm}.
\end{equation}

Taking for this mass ratio the mean radius of the Roche lobe to be
$R_L \approx 0.5 a$, the tidal truncation radius, i.e. disc
radius, is $R_d \approx 0.8 R_L \approx 0.4 a$. Using this, and the
mass ratio $q = M_2/M_1 = 0.25$ we find (Bate et al., 2000) that for
this system (again assuming a power--law surface density
$\Sigma \propto R^{-3/2}$)
\begin{equation}
\frac{\omega_p}{\Omega_d} \sim 0.005.
\end{equation}
Then with $\alpha_h \approx 0.2$ and $H/R \approx 0.03$ we find
from equation(\ref{talignvisc}) that during outburst the disc
alignment timescale is
\begin{equation}
t_{\rm align}({\rm outburst}) \sim 0.92 \omega_p^{-1} \sim 0.42 {\rm d}.
\end{equation}
Thus it looks as if maintaining the disc tilt through normal outbursts
is problematic. We are therefore left with the necessity that the disc
tilting mechanism operates also during ordinary outbursts in this
system.

Thus in any case, we require the disc tilt to last a least throughout
the 30 d period between ordinary outbursts, i.e. for 385 orbital
periods, or 10 precession periods, and we also need to the disc tilt to
be enhanced or at least maintained during normal outbursts.

The problem now is that during quiescence, in the standard models,
$(H/R)^2$ {\em decreases} by a factor of $\sim 10 - 25$, which,
although $\alpha$ decreases by of order $\sim 10$, implies that
$t_{\rm align}$ is just as small, or even smaller, in quiescence.

\subsection{Is $\alpha$ even lower in quiescence?}

This implies that we need to reduce $\alpha$ during quiescence below
the value usually assumed for $\alpha_c$. But doing this has a
significant consequence: it implies that we move into the regime where
$\alpha < H/R$, and warp propagation occurs in a wavelike manner
through pressure effects.

In this regime, the disc can precess as a solid body provided that
(Bate et al 2000) $R_d/c_s \le \Omega_p^{-1}$, or, provided that
\begin{equation}
H/R > \omega_p/\Omega_d \approx 0.005,
\end{equation}
which is approximately satisfied.

Using $H/R \approx 0.006$ we find $t_{\rm align} \sim 0.34
(\alpha/0.01)^{-1}$~d. This implies that to get $t_{\rm align}
\ge 30$~d we would require $\alpha \le 10^{-4}$.

\section{Conclusions}

\subsection{Her X-1/HZ Her}

The Her X--1/HZ Her system is roughly consistent with the simple
Navier-Stokes viscosity approximation for modelling the evolution of
disc tilt (Ogilvie 1999), provided that $\alpha \approx 0.3$ which is
what is expected for a fully ionized disc, although the rough
estimates indicate that the disc alignment timescale might be too
small by a factor of a few. More accurate modelling is required before
a definitive conclusion can be drawn.

This illustrates the importance of observing the behaviour of Her X--1
during its anomalous low states, when the disc appears to align more
closely with the orbital plane. Even more important will be to observe
the system through a prolonged low state such as that reported by
Jones, Forman \& Liller (1973), when mass transfer seems to have
halted altogether in this system for a period of around 10 years.

\subsection{V503 Cyg}

In the dwarf nova V503 Cyg, to maintain the amplitude of the negative
superhump throughout the outburst cycle -- that is from superoutburst
to superoutburst, so for $\simeq 88$~d -- it is necessary that the
mechanism which produces the disc tilt should be present during
ordinary outbursts as well as superoutbursts. Otherwise the tilt is
expected to decay through an ordinary outburst.

For the negative superhump to survive throughout the interoutburst
period (of order 10 precession periods) requires that $\alpha \le
10^{-4}$, a value much smaller than the value $\alpha_c \sim 0.01$
usually assumed during quiescence.

Indeed, $\alpha$ has to be sufficiently small that we are in the
regime in which disc warp is propagated in a wavelike manner. For
typical numbers, the wave crossing timescale for the disc is
comparable to the precession period, so that the disc close to the
limit at which is might be torn into segments.

\section{Discussion}

Current thinking is that `viscosity' results from MHD turbulence,
using feedback via the magnetorotational instability (MRI). King et
al. (2007) noted that for fully ionized discs, numerical modelling of
this type does not appear to produce agreement with the observations.
Most computational work involves shearing boxes, and seems to indicate
that in the absence of a large-scale external magnetic field, the disc
itself cannot provide a large enough viscosity (e.g. Bai \& Stone
2012) by at least an order of magnitude.

MRI-driven MHD turbulence derives its energy from radial angular
momentum transfer, so it is unclear how relevant this is to the
problem of warped discs considered here.  As Pringle (1992) has
remarked there is a radical difference in the reaction of an embedded
magnetic field to a continuous $R\phi$ shear, compared to the
oscillatory $Rz$ shear which results from a disc warp. So it is
important to ask how well models of disc behaviour based on
Navier-Stokes viscosity (small mean free path, isotropic) manage
to account for the evolution of disc warp.

We have seen that for fully ionized discs (such as that in Her X-1 and
in dwarf novae in outburst) this theory seems to work adequately.  We
note that the initial rough estimates made here suggest that
theoretical alignment timescales might be shorter than those observed
by factors of a few. This could of course mean that the force causing
the misalignment does not at first switch off completely, and slows
the realignment beyond the viscous estimate. This seems unlikely for
Her X-1, which is close to the stability boundary for radiation
warping as we noted above. It is possible that some other mechanism,
such as the effect of the neutron star magnetic field (e.g. Lai, 1999;
Terquem \& Papaloizou, 2000; Murray et al., 2002) might maintain a
residual tilt and slow the alignment, although this is unlikely for
the outer disc. More detailed calculations are required before we can
be sure of this discrepancy. This offers a challenge to MHD modelling:
to what extent can this theory be made to agree with observations,
both the 35~d--period behaviour and the transition to anomalous low
states. As King et al. (2007) emphasized, even to explain $\alpha
\approx 0.3$ is likely to require taking the computations beyond
shearing boxes towards more global simulations encompassing the whole
disc (unless all discs have exactly the right amount of external
magnetic field threading them). It is exactly such 3D global models
which will be required to model the evolution of disc warps.

In quiescent dwarf nova discs the nature of the viscosity is unclear,
since these discs are cool enough that the usual models of MRI and MHD
turbulence are unlikely to apply (Gammie \& Menou, 1998). Using the
Navier-Stokes viscosity model we have noted that although all models
of dwarf nova outbursts seem to demand that $\alpha_c \sim 0.01$, the
value of $\alpha$ required to allow negative superhumps to persist
throughout the period between outbursts is smaller than this by at
least two orders of magnitude. Thus either it is possible to produce
models of DN outbursts with lower $\alpha_c$, or the Navier-Stokes
model is inadequate.

This leaves us with two questions.

(i) does the `viscosity' (whatever it is) act like a Navier-Stokes
viscosity?

(ii) is it possible to develop numerical models that work, not just
for fully ionized, strongly conducting discs, but also for low
temperature, low conductivity discs?

Besides these two theoretical questions, it is clear that further
observational work constraining the behaviour of warped and tilted
discs would be very valuable.

\section*{Acknowledgments}

JEP thanks STScI for support from their Visitor Program. SHL thanks
IoA, Cambridge, for support from their Visitor Programme.

\label{lastpage}
\end{document}